\documentclass[journal=jpccck,manuscript=letter]{achemso}
\setkeys{acs}{articletitle = true}
\setkeys{acs}{etalmode=truncate}
\setkeys{acs}{maxauthors=10}
\usepackage{booktabs}
\usepackage{tabularx}
\usepackage{xr}
\usepackage[version=3]{mhchem} 
\usepackage{textcomp}
\usepackage{xcolor}
\usepackage{soul}
\usepackage[normalem]{ulem}

\newcommand{\ReS}{\mathrm{ReS}_2}

\newcommand{\MoS}{\mathrm{MoS}_2}
\newcommand{\WS}{\mathrm{WS}_2}

\newcommand{\DA}{\Delta A}
\newcommand{\EX}{E_{\mathrm{b,X}}}
\newcommand{\Eg}{E_{\mathrm{g}}}
\newcommand{\Eopt}{E_{\mathrm{g,o}}}
\newcommand{\A}{A_{0}}
\newcommand{\DE}{\Delta E}
\newcommand{\DS}{\Delta S}
\newcommand{\DW}{\Delta w}

\author{Pieter Schiettecatte}
\affiliation{Physics and Chemistry of Nanostructures, Department of Chemistry, Ghent University, 9000 Ghent, Belgium}
\alsoaffiliation{Center for Nano and Biophotonics, Ghent University, 9000 Ghent, Belgium}
\email{pjschiet.schiettecatte@ugent.be}

\author{Zeger Hens}
\affiliation{Physics and Chemistry of Nanostructures, Department of Chemistry, Ghent University, 9000 Ghent, Belgium}
\alsoaffiliation{Center for Nano and Biophotonics, Ghent University, 9000 Ghent, Belgium}

\author{Pieter Geiregat}
\affiliation{Physics and Chemistry of Nanostructures, Department of Chemistry, Ghent University, 9000 Ghent, Belgium}
\alsoaffiliation{Center for Nano and Biophotonics, Ghent University, 9000 Ghent, Belgium}
\email{Pieter.Geiregat@UGent.be}

\title[]
  {A Roadmap to Decipher Ultrafast Photophysics in Two-Dimensional Nanomaterials}


\begin{document}

\begin{tocentry}
\begin{center}
	\includegraphics[scale=0.7]{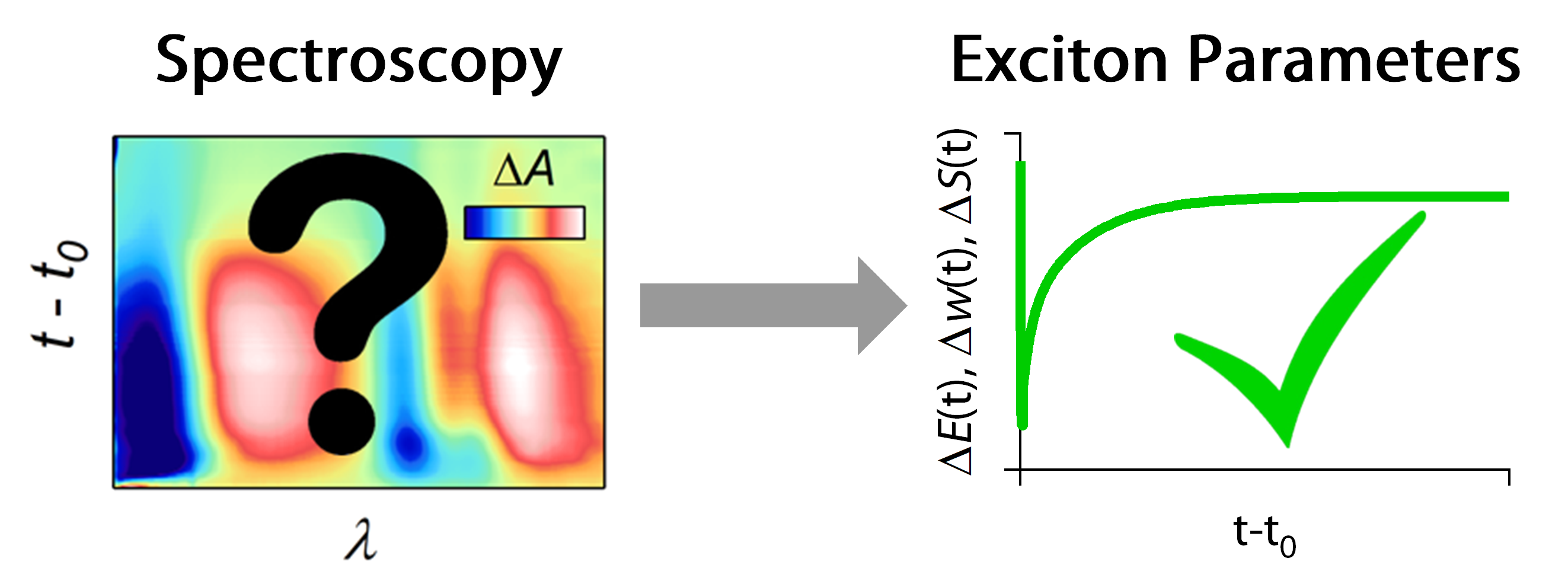}
\end{center}
\end{tocentry}


\begin{abstract}

Atomically thin two-dimensional (2D) semiconductors are extensively investigated for opto-electronic applications that require strong light-matter interactions. In view of such applications, it is essential to understand how (photo)excitation alters the non-linear optical response of these materials under high carrier density conditions. Broadband transient absorption (TA) spectroscopy is by now a widely used tool to study semiconductor physics in such highly excited systems. However, the complex interplay between different many-body interactions in 2D materials produces highly congested spectral information and an ensuing non-trivial nonlinear photo-response, thereby masking the desired intrinsic photophysics. Herein, we outline a concise roadmap for analyzing such congested datasets based on examples of TA analysis of various 2D materials. In particular, we emphasize the synergy between an initial qualitative understanding of the transient photo-response based on line shapes and their derivatives, and a consequent quantitative spectral deconvolution backed by such insights. 

\end{abstract}


\section{Introduction}

\begin{figure}
\centering
\includegraphics{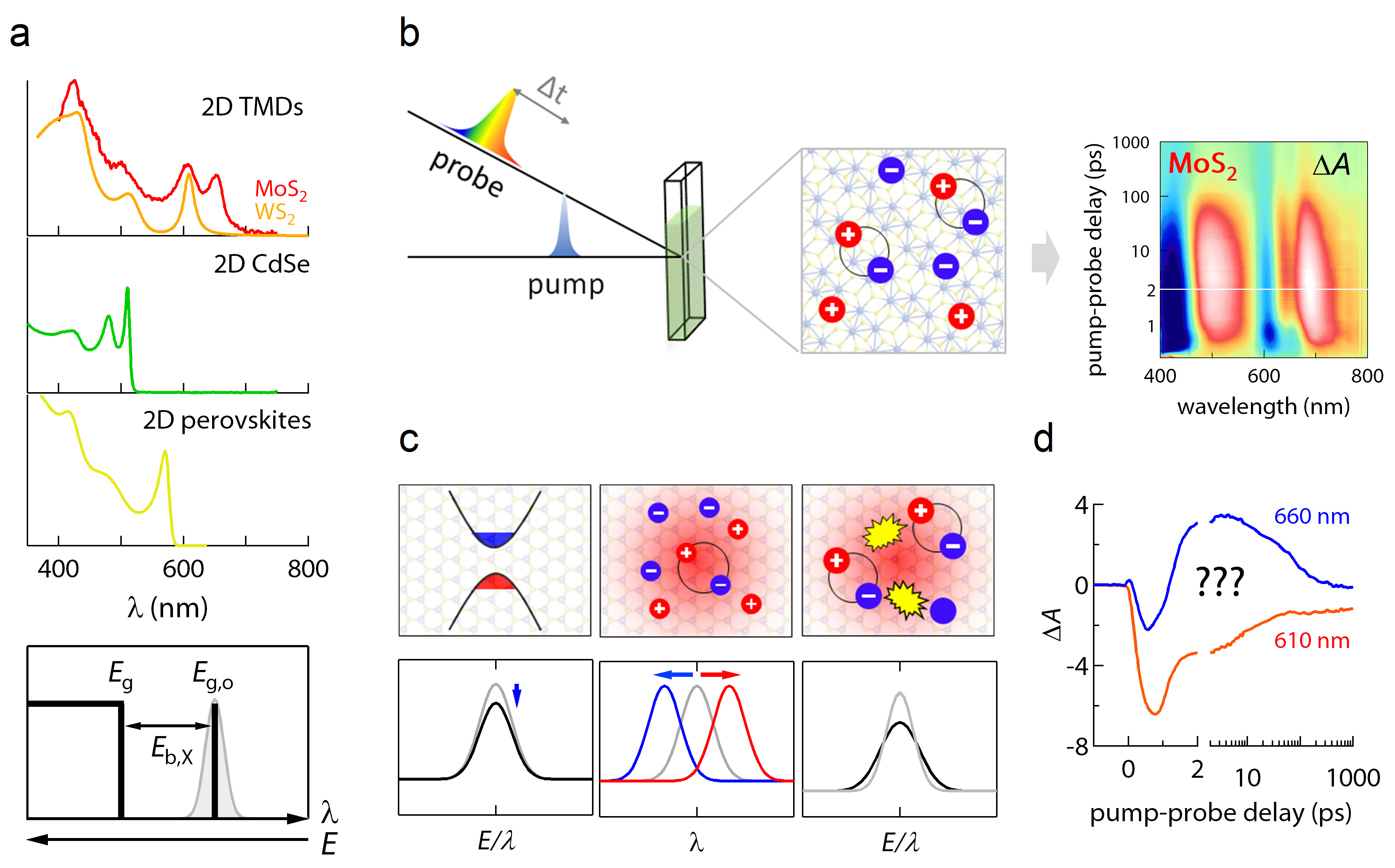}
\caption{(a) (top) Exemplary optical absorbance spectra of two-dimensional semiconductors including (from top to bottom) $\MoS$ and $\WS$,\cite{li2014measurement} 2D $\mathrm{CdSe}$ nanoplatelets,\cite{di2020colloidal} and 2D hybrid organic inorganic perovskites.\cite{song2020determination} The optical absorbance is characterized by a sharp exciton line superimposed on a 2D DOS-like continuum. (bottom) Schematic depicting the absorbance spectrum of a two-dimensional excitonic semiconductor with an optical band gap $\Eopt$ and an exciton transition at binding energy $\EX$ superimposed on a step-like continuum at energy $E_g$. (b) Schematic depicting the principles behind ultrafast pump-probe spectroscopy. A specimen is pumped with an ultrafast laser pulse, inducing changes in the absorbance. A delayed probe pulse monitors changes in absorbance. Generally, experiments are mapped by tracking changes in the differential absorbance ($\DA$) as a function of (ordinate) pump probe delay and (abscissa) wavelength/energy, of which an example color map is shown for colloidal MoS$_2$. Adapted with permission from Reference~\citenum{schiettecatte2019ultrafast}. Copyright 2019 American Chemical Society. (c) Pictorial representation of the most common transient effects after photo-excitation, including (from left to right) changes in the integrated absorbance brought by state filling and Coulomb screening, shifts in the transition energy due to band gap renormalization and binding energy reduction, and linewidth changes induced by collisional broadening through carrier-carrier and carrier-phonon scattering. (d) Spectral slices of $\DA$ evaluated at the $A$ (660 nm) and $B$ (610 nm) transition energies, extracted from the color map in panel c, highlighting the difficulty to extract reliable information based on slices at a fixed probe wavelength.}\label{Fig_Intro}
\end{figure}

Advancements in synthetic chemistry and material characterization over the last decades have prompted the discovery of strongly confined two-dimensional (2D) semiconductors with atomically thin dimensions. The most notable examples include the large family of transition metal di-chalcogenides (TMDs), cadmium chalcogenide nanoplatelets, and the recently emerging library of layered 2D perovskites (Figure~\ref{Fig_Intro}a).\cite{mak2010atomically,splendiani2010emerging,ithurria2011colloidal,shi2018two} In all these examples, a strong out-of-plane carrier confinement alongside a reduced dielectric screening of the Coulomb interaction enhances many-body interactions between photo-excited charge carriers. As a consequence, otherwise unbound electrons and holes condense into tightly bound excitons with unparalleled exciton binding energies at room temperature.\cite{wang2018colloquium,naeem2015giant,ugeda2014giant,wang2018quantum} At higher densities, these excitons can further condense into bi-excitonic molecules, as observed in several 2D materials, even at room temperature. Compared to continuum transitions involving unbound carriers, light-matter interaction at the exciton transition energy is greatly enhanced, causing a pronounced absorption of light with large optical absorption coefficients in the order of $\mathrm{10^6\:cm^{-1}}$.\cite{mak2010atomically,bernardi2013extraordinary} In summary, the envisaged prospect of strong absorption and, in parallel, efficient emission through excitons at room temperature, alongside the tantalizing possibility to explore multi-exciton physics, is what stimulates research into these atomically thin semiconductors.\cite{chen2014quasi,grim2014continuous,ye2015monolayer,ross2014electrically,cheng2014electroluminescence}

The study of the underlying ultrafast phenomena dictating exciton formation and their dissociation or association in strongly excited systems heavily relies on various forms of pump-probe spectroscopy, \textit{e.g.}, the increasingly prevalent transient absorption (TA) spectroscopy where visible/near-infrared light pulses are used as pump and/or probe. Through these methods exciton properties can be disentangled at otherwise inaccessible time and energy scales (Figure~\ref{Fig_Intro}c).\cite{maiuri2019ultrafast} Focusing on 2D nanomaterials, initial pump-probe studies were often limited to single color/wavelength probing due to the technological limitations of measuring single flakes with broadband probes using microscopy.\cite{wang2014ultrafast,wang2012ultrafast,sun2014observation} Even though such an approach captures the charge carrier dynamics, looking to the spectrally strongly varying linear and transient absorption in Figures 1a and 1b, one can understand that this often impedes a consistent interpretation. Lately, however, the development of scalable production methods -- liquid-phase exfoliation (LPE), colloidal synthesis, and large area chemical vapor deposition (CVD) -- together with increasing availability of TA microscopy equipment has resulted in a shift toward broadband TA spectroscopy.\cite{zhou2019ultrafast,chernikov2015population,schiettecatte2019ultrafast,cunningham2016charge,pogna2016photo,sie2017observation,cunningham2017photoinduced,sim2013exciton,ruppert2017role,calati2021ultrafast,trovatello2022disentangling,aivazian2017many,schiettecatte2021unraveling,calati2022dynamic,taank2022insightful,karmakar2021electron,bera2021atomlike} 

Such broadband TA studies highlighted that the photophysics of 2D materials is far from trivial to interpret based on spectral information. Figure~\ref{Fig_Intro}b depicts an example of a 2D time-wavelength map obtained after photo-excitation of MoS$_{2}$, a typical 2D material, featuring a complex interplay between spectral shifts, linewidth changes and changes in the integrated absorbance of the exciton resonances. These effects, summarized in Figure 1c, on the line shape and probability of transitions are generally rationalized based on the many-body theory of quasi-2D quantum wells (QWs) which predicts drastic changes in the optical absorption of a QW when additional (free) carriers are present in the system.\cite{schmitt1989linear,haug2009quantum} Most notably, photo-excited carriers are predicted to modify the optical band gap through band gap renormalization (BGR) and exciton binding energy reduction (BER), broaden the exciton linewidth due to collisional broadening, and reduce its oscillator strength through state filling (Pauli blocking) and Coulomb screening of the \textit{e-h} attraction (Figure~\ref{Fig_Intro}c). As these effects typically occur simultaneously, the spectral information is often highly congested and non-trivial to interpret, and kinetic traces can widely differ depending on which probe wavelength they are evaluated at (Figure~\ref{Fig_Intro}d). In particular, these concerns raise questions on how exciton parameters can be reliably extracted from such spectrally complex $\DA$ maps.

Herein, we outline a concise roadmap for analyzing spectrally complex transient absorbance ($\DA$) maps based on recent literature examples. The eventual goal is to obtain a rational methodology to go from congested 2D time-wavelength maps to clear kinetic information on exciton parameters as the exciton energy change $\Delta E$, the linewidth change $\Delta w$ and the change of oscillator strength or the integrated change in absorbance $\Delta S$. To this end, we first develop a series of qualitative insights for each observable into the shape of transient spectra based on a derivative analysis of the well-understood linear absorption spectrum. In particular, we develop this approach for spectral shifts ($\Delta E$) and linewidth changes ($\Delta w$).  Based on these insights, we can more rationally optimize global fitting procedures to attain the net $\Delta S$, often directly interpreted as the 'carrier dynamics'. 

\subsection{Spectral Shifts}

\begin{figure}
\centering
\includegraphics{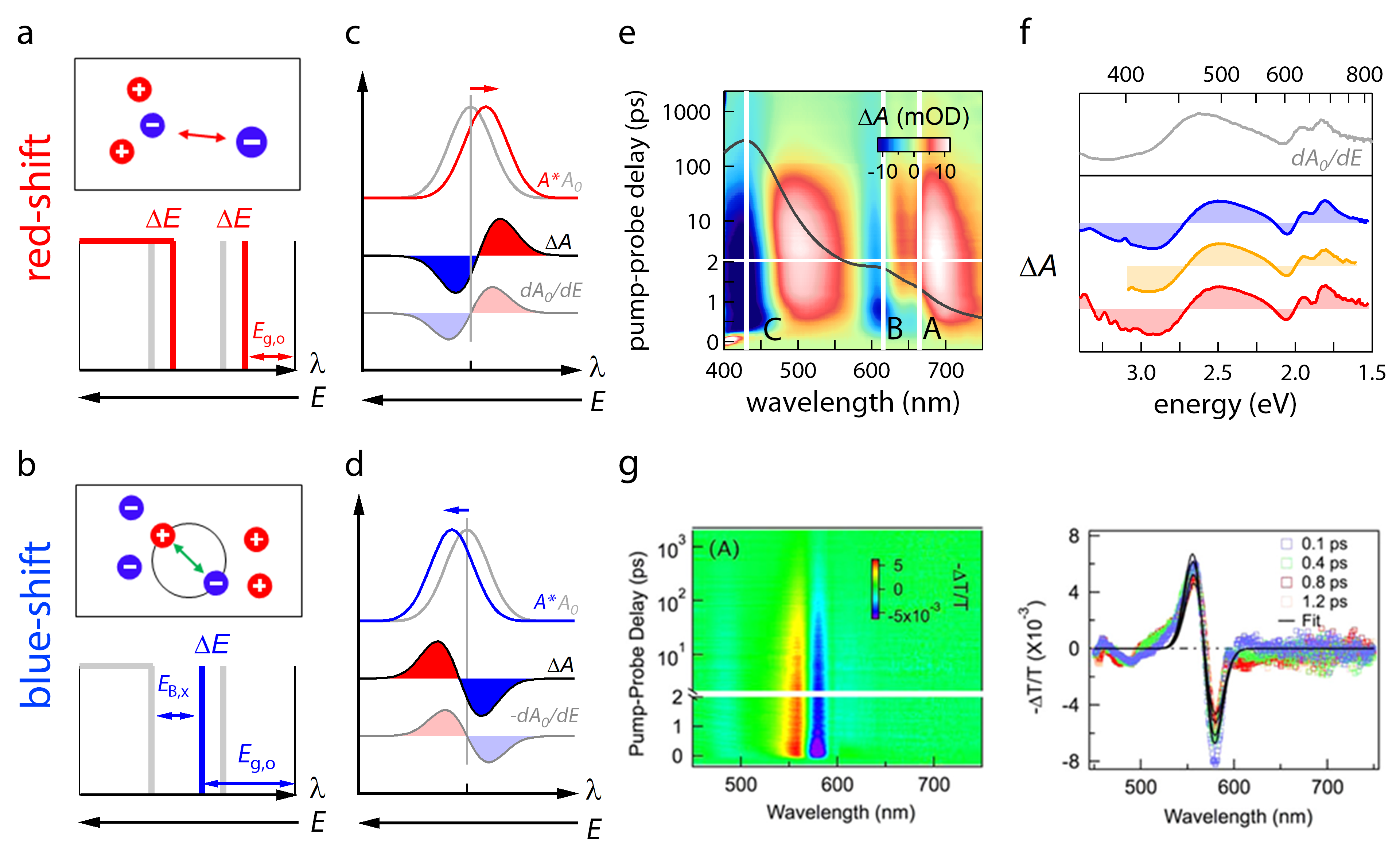}
\caption{Schematics depicting (a) a spectral red-shift due to band gap renormalization and (b) a spectral blue-shift due to binding energy renormalization. $E_{\mathrm{g,o}}$, $E_{\mathrm{B,X}}$ and $\Delta$\textit{E} denote the optical band gap, the exciton binding energy and the spectral shift, respectively. (c,d) Schematics depicting the effect of a spectral (c) red-shift and (d) blue-shift on the transient absorbance $\DA$. A spectral shift results in a $\DA$ spectrum resembling a first-derivative-like spectrum $dA_{0}/dE$. (e) The transient response of colloidal molybdenum disulfide nanosheets as an exemplary color map of a red-shift.\cite{schiettecatte2019ultrafast} Particularly at longer time delays, $\DA$ shows multiple bands of bleach and PIA mirrored around the $C$-band, $B$ and $A$ exciton. Adapted with permission from Reference~\citenum{schiettecatte2019ultrafast}. Copyright 2019 American Chemical Society. (f) Traces of $\DA$ probed at different excitation energies, extracted from color maps similar to those shown in panel e, together with the first derivative of the linear absorbance to the energy (grey) $dA_{0}/dE$. The spectral shape of $\DA$ is independent on the excitation energy and resembles $dA_{0}/dE$, as is indicative of a spectral red-shift of all transition energies due to BGR dominating over BER. (g) The transient response of bilayer lead iodide perovskites as a didactic color map and spectral slice showing a clear blue-shift of the exciton transition.\cite{wu2015excitonic} $\DA$ shows an alternate sequence of PIA and bleach mirrored around the $1S$ exciton. Reprinted with permission from Reference~\citenum{wu2015excitonic}. Copyright 2015 American Chemical Society.}\label{Fig_Shifts}
\end{figure}

As depicted in Figure~\ref{Fig_Intro}a, the optical band gap $E_{\mathrm{g,o}}$ of an excitonic semiconductor depends on the free particle band gap $E_g$ and the exciton binding energy $E_{\mathrm{b,X}}$, such that $\Eopt$ equals $\Eg - \EX $.\cite{wang2018colloquium} Compared to an unexcited semiconductor, carrier creation and/or subsequent heat generation through photo-excitation alters the free particle band gap and the exciton binding energy.\cite{schmitt1986many} A photo-induced shrinkage of $E_g$, referred to as band gap renormalization, results from screening of repulsive Coulomb and exchange-correlation interactions.\cite{kalt1992band} Similarly, photo-induced changes in lattice temperature and non-equilibrium phonon populations after cooling of hot carriers affect the free particle band gap, where typically a higher lattice temperature decreases $E_g$.\cite{ruppert2017role} As shown in Figure~\ref{Fig_Shifts}a, both effects manifest themselves as a shrinkage of the optical band gap, which concomitantly red-shifts transition energies. Additionally, in excitonic systems, photo-excited \textit{e-h} pairs dynamically screen (or reduce) the electron-hole attraction that binds the exciton, thereby reducing the exciton binding energy and reducing its oscillator strength. As illustrated in Figure~\ref{Fig_Shifts}b, binding energy reduction (BER) shifts the exciton absorption toward higher energy (blue-shift). Both mechanisms work in spectrally opposite directions, yet typically do not fully compensate in quasi-two dimensional systems. \cite{wake1992response} As such, the direction of spectral shifts observed experimentally depends on an intricate balance between BER and BGR.\cite{steinhoff2014influence} 

\paragraph{Effect on $\Delta A$ and Derivative Analysis} Figures~\ref{Fig_Shifts}c and~\ref{Fig_Shifts}d illustrate the effect of a change in transition energy on the transient absorbance $\DA$ through the shift of an isolated Gaussian absorption band to lower energy (red-shift) and higher energy (blue-shift), respectively. Clearly, such a spectral shift reduces the absorbance at one side of the absorption band, while increasing it at the other side, without changing the integrated absorbance. A spectral shift therefore yields a $\DA$ spectrum that resembles the first derivative of the absorbance spectrum, consisting of an alternated sequence between bleach (negative $\DA$, blue) and photo-induced absorption (PIA, positive $\DA$, red) for the example shown. Note that the specific bleach/PIA sequence depends on the direction of the transition's energy shift.

A particularly useful approach to further analyse spectral shifts involves combining such a conceptual interpretation with an analysis based on the derivatives of the absorbance spectrum given that, for small enough shifts, the ratio of $\DA$ and $dA_{0}/dE$ is directly proportional to the spectral shift $\Delta E$ according to:\cite{schiettecatte2019ultrafast,geiregat2014coulomb}
\begin{equation}
\Delta E \approx \frac{\Delta A}{(dA_{0}/dE)}
\label{TAS_MoS:Eq SpectralShift}
\end{equation}
We illustrate this point in Figure~\ref{Fig_Shifts}f by plotting traces of $\DA$, extracted from similar color maps to the one shown for $\MoS$ in Figure~\ref{Fig_Shifts}e, together with $dA_{0}/dE$. Clearly, features in the TA spectrum coincide with local minima and maxima in $dA_{0}/dE$. From Equation~\ref{TAS_MoS:Eq SpectralShift}, we could further estimate the magnitude of the spectral red-shift, which in this case amounted to $\approx\mathrm{20\:meV}$.\cite{schiettecatte2019ultrafast} Importantly, this number closely matched values extracted from a spectral deconvolution through a global fit (see Figure~\ref{Fig_Fit}), making this approach valuable to validate numerical fitting approaches.

\paragraph{Literature Examples} Throughout literature, spectral red-shifts have been reported for various transition metal di-chalcogenides produced by exfoliation, colloidal synthesis and chemical vapor deposition.\cite{zhou2019ultrafast,bera2021atomlike,cunningham2017photoinduced,chernikov2015population,schiettecatte2019ultrafast,cunningham2016charge,pogna2016photo,sie2017observation,karmakar2021electron} Often, those spectral red-shifts were accompanied by an exciton linewidth broadening,\cite{chernikov2015population,zhou2019ultrafast,cunningham2016charge,schiettecatte2019ultrafast,sie2017observation,ruppert2017role} see further. The color map recorded on few-layer colloidally dispersed molybdenum disulfide (\ce{MoS_2}) in Figure~\ref{Fig_Shifts}e is a typical example of such a red-shifted $\DA$ spectrum.\cite{schiettecatte2019ultrafast} Particularly at time delays beyond $10\:\mathrm{ps}$, the transient absorbance features a consecutive sequence of bleach and PIA bands centered around the excitonic resonances, which reflects band gap renormalization.\cite{pogna2016photo} Similarly, blue-shifted exciton transitions have been observed in quantum wells\cite{wake1992response,schultheis1986ultrafast,leite1969effect,hulin1986well}, atomically-thin TMDs,\cite{cunningham2017photoinduced,sim2013exciton,ruppert2017role,calati2021ultrafast,trovatello2022disentangling} and 2D perovskites.\cite{wu2015excitonic} In $\MoS$\cite{sim2013exciton} and $\mathrm{WS_2}$\cite{cunningham2017photoinduced,ruppert2017role,calati2021ultrafast,trovatello2022disentangling}, blue-shifts at early delays are typically accompanied by a broadening of the exciton band due to carrier scattering, see further section on broadening. Cross-overs from blue-to-red-shifts have been observed in $\mathrm{WS_2}$,$\mathrm{WSe_2}$, and $\mathrm{ReS_2}$, evidencing the intricate balance between BGR and BER on the excitation density.\cite{aivazian2017many,cunningham2017photoinduced,calati2021ultrafast,schiettecatte2021unraveling,calati2022dynamic} A revealing case of a relatively uncongested spectral blue-shift is provided in Figure~\ref{Fig_Shifts}g.\cite{wu2015excitonic} Here, the $\DA$ map of a bilayer lead iodide perovskite shows a first-derivative-like signature, mirrored around the fundamental exciton at 570 nm, with the PIA band being at the high energy side and bleach band at the lower energy side of the exciton line. Take note that the sequence of PIA/bleach is opposite from the one shown in Figures~\ref{Fig_Shifts}e-f and thus characterizes a blue-shift.

\subsection{Linewidth Changes}

\begin{figure}
\centering
\includegraphics{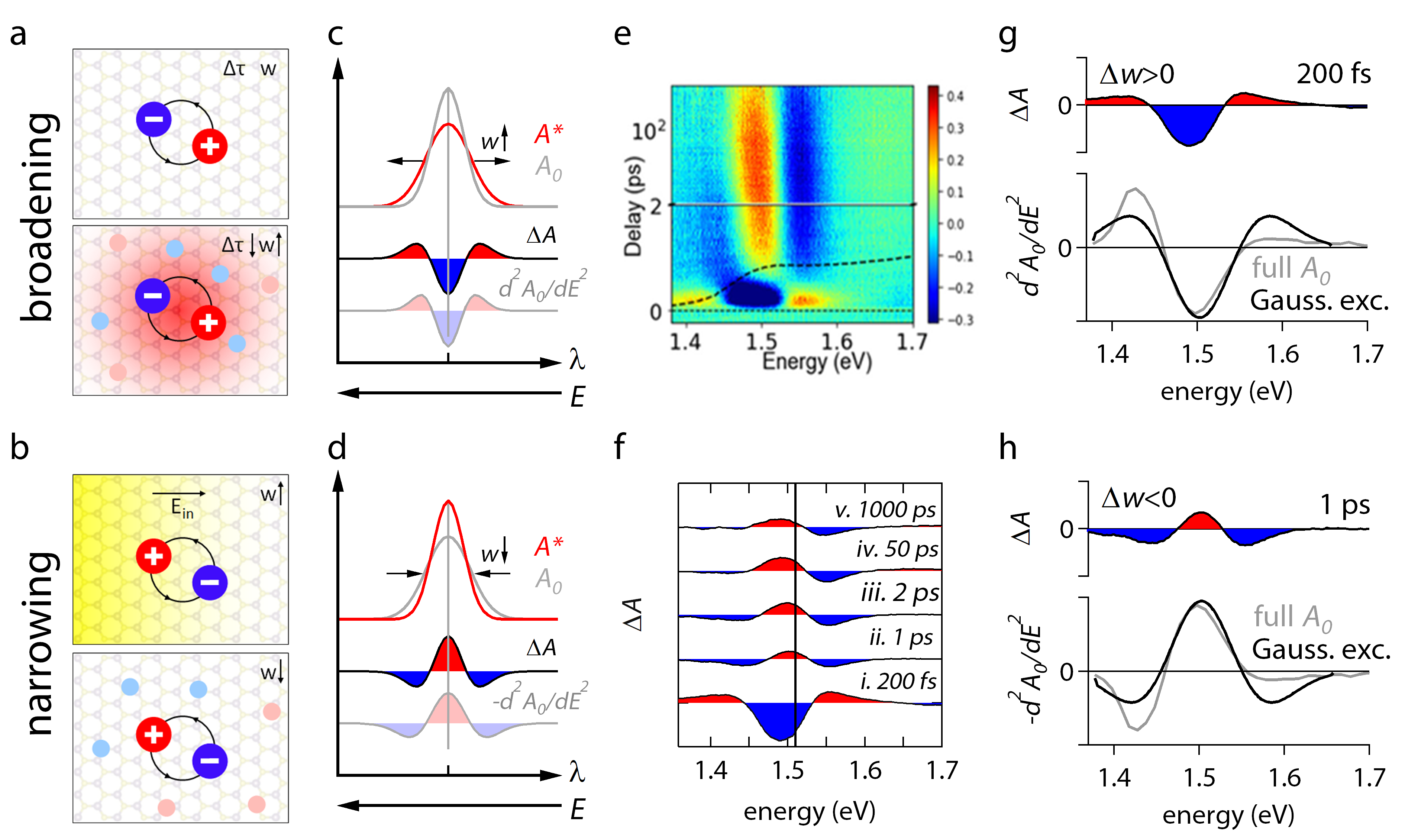}
\caption{Schematics depicting (a) linewidth broadening $\Delta w>0$ due to collisional broadening and (b) linewidth narrowing $\Delta w>0$ due to screening of an internal electric field. $w$ and $\Delta \tau$ respectively denote the width of the exciton band, and the exciton de-phasing time. (c,d) Schematics depicting the effect of a linewidth (c) broadening and (d) narrowing on the transient absorbance $\DA$. Linewidth changes lead to $\DA$ spectra that resemble the second derivative of the linear absorbance $\A$ to the energy $E$ ($d^2 A_{0}/d E^2$). (e) Overview of the transient absorbance $\DA$ recorded on a film of few-layer $\ReS$, showing a complex interplay between several spectral effects. The dashed line represents the linear absorbance spectrum $A_0$. (f) Traces of the $\DA$ map in panel e sliced at specific pump-probe time delays, highlighting the evolution of (i) a line broadening compounded by a blue-shift into (ii-iii) a line narrowing (iv-v) congested by a red-shift at an even longer delay. (g,h) Overlay plots of $\DA$ with $d^2 A_{0}/d E^2$, highlighting a one-to-one correspondence of a broadened/narrowed $\DA$ spectrum with a second derivative spectrum. The grey and black lines represent the derivatives of $A_0$ and the isolated exciton profile, respectively. Reprinted and adapted with permission from Reference~\citenum{schiettecatte2021unraveling}. Copyright 2021 American Chemical Society.}\label{Fig_Width}
\end{figure}

The homogeneous linewidth of an exciton transition is determined by the de-phasing rate or coherence time of the exciton state(s).\cite{moody2015intrinsic} A loss of exciton coherence can occur due to (in)elastic scattering of excitons with carriers or lattice vibrations, generally referred to as carrier-carrier and carrier-phonon scattering. Collisional broadening due to an increase in carrier population or lattice temperature is therefore the dominant broadening mechanism of absorption (and emission) line(s) in semiconductors, see Figure~\ref{Fig_Width}a.\cite{wake1992response,sim2013exciton} In reality, transitions can also inhomogeneously broaden due to local potentials arising from defects.\cite{stoneham1969shapes,raja2019dielectric} While being less prominent in the literature, photo-induced line narrowing has been reported as a result of carrier-induced screening of parasitic local defect-induced potentials or externally applied and built-in electric fields, see Figure~\ref{Fig_Width}b.\cite{miller1985electric,schmitt1989linear,schiettecatte2021unraveling,shikanai2000pump,chen2003ultrafast,chen2003excitonic,omae2003effects,omae2002effects}

\paragraph{Effect on $\Delta A$ and Derivative Analysis} Figures~\ref{Fig_Width}c and~\ref{Fig_Width}d illustrate the effect of a linewidth change on the transient absorbance $\DA$ by, respectively, broadening and narrowing an isolated Gaussian absorption band. Clearly, broadening (narrowing) a resonance decreases (increases) the absorbance maximum, while increasing (decreasing) the intensity at the tails of the absorbance band. A central bleach sided by two positive side bands of photoinduced absorption therefore reflects line broadening, whereas the pattern is reversed for narrowed lines. Generally, a linewidth change can thus be best described as a second-derivative-like $\DA$ spectrum. For Gaussian line shapes and small enough width changes, the ratio of $\DA$ and $d^2 A_{0}/d E^2$ is directly proportional to the change in width $\Delta\sigma=\sigma-\sigma_0$ according to:\cite{schiettecatte2021unraveling}
\begin{equation}
\Delta \sigma \approx \frac{\Delta A}{(d^2 A_{0}/d E^2)\sigma_0}
\label{TAS_MoS:Eq SpectralWidth}
\end{equation}

\paragraph{Literature Examples} The $\DA$ map of liquid-phase exfoliated (LPE) $\ReS$ in Figure~\ref{Fig_Width}e serves to illustrate broadened and narrowed resonances, and their relation to the second derivative of the linear absorbance to the energy.\cite{schiettecatte2020liquid,schiettecatte2021unraveling} A more detailed description of the peculiar photophysics observed in LPE $\ReS$ can be found in Ref.~\citenum{schiettecatte2021unraveling}. To analyse this map conceptually, we present traces of $\DA$ at specific pump-probe time delays in Figure~\ref{Fig_Width}f and overlay plots of $\DA$ with $d^2 A_{0}/d E^2$ in Figures~\ref{Fig_Width}g-h. Shortly after photoexcitation ($t$=200 fs, spectrum i), the transient resembles a second derivative-like spectrum (see Figure~\ref{Fig_Width}g), and therefore predominantly reflects a broadened absorption profile. Notwithstanding the overlap of $\DA$ with $d^2 A_{0}/d E^2$, the bleach center of $\DA$ is slightly off-set toward lower energy (see vertical line), likely due to a superimposed blue-shift. From 1 ps onwards (spectra ii and iii), the transient has inverted and mimics $-d^2 A_{0}/d E^2$ (see Figure~\ref{Fig_Width}h), and thus reflects a narrowed resonance. At longer time delays (spectra iv-v), a red-shift grows in, resulting in a transient with a first derivative-like shape. This example points out how a conceptual interpretation of the color map based on simple line shapes, in addition to comparing slices of $\DA$ to first and second derivatives, can greatly contribute to decrypt TA maps qualitatively, prior to attempting more elaborate curve fits.

\subsection{Extracting Exciton Parameters through Curve Fits}

\begin{figure}
\centering
\includegraphics{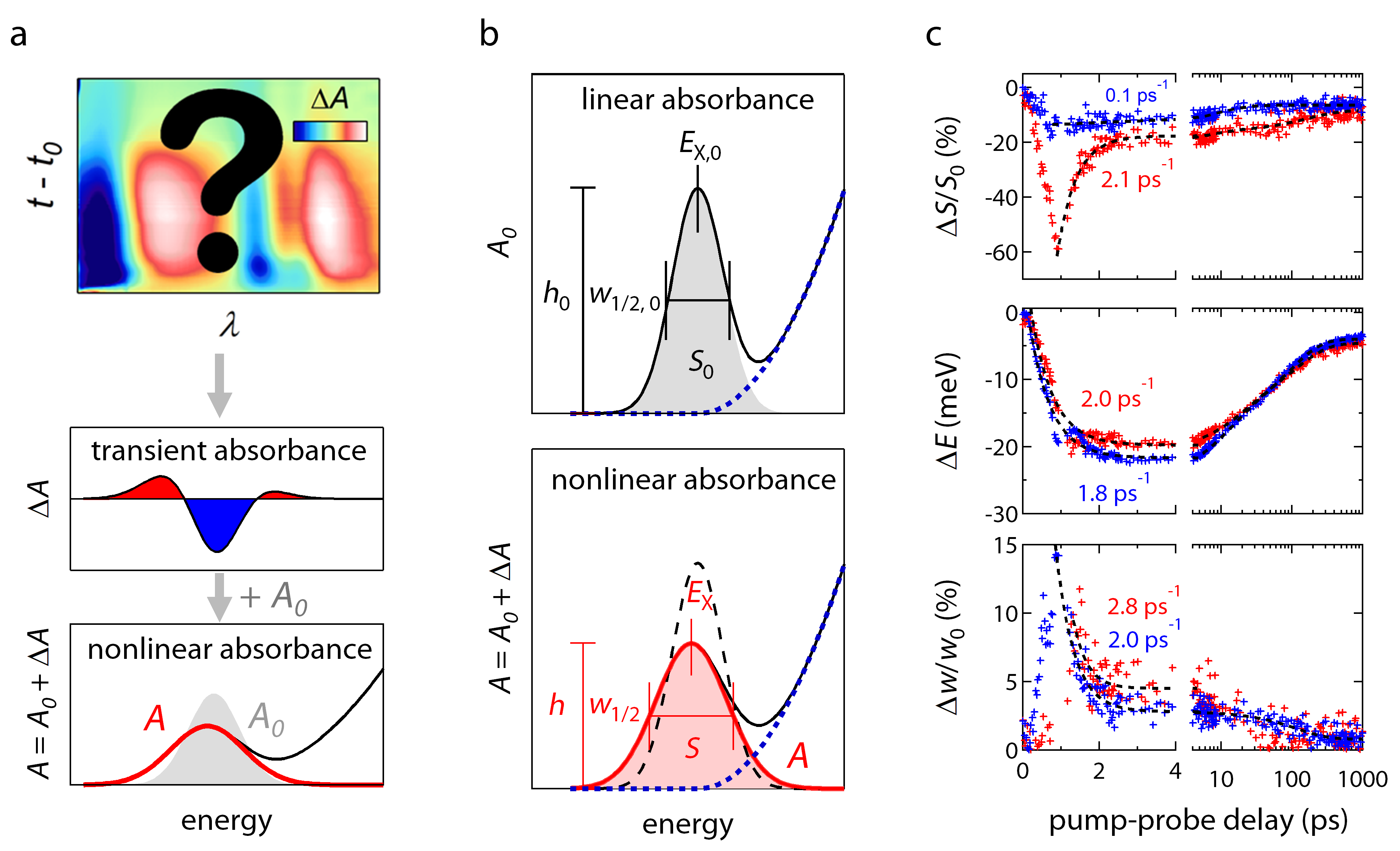}
\caption{Overview of the general principle behind a spectral deconvolution of a $\DA$ map. (a) The transient absorbance $\DA$ is summed together with the linear absorbance $\A$, yielding the nonlinear absorbance $A$. (b) $\A$ and $A$ are fit to (filled curve) a Gaussian function and (blue dashed line) a polynomial background, yielding the central position $E_{X,(0)}$, the area $S_{(0)}$ and the width $w_{1/2,(0)}$ of the exciton band. (c) Repeating step b as a function of pump-probe delay maps the time-dependent evolution of these fit parameters. This is illustrated by fitting the $\DA$ map of colloidal $\MoS$ in Figure~\ref{Fig_Shifts}e to a sum of two Gaussians, accounting for the (red markers) $A$ and (blue markers) $B$ exciton, and a polynomial background.\cite{schiettecatte2019ultrafast} Adapted with permission from Reference~\citenum{schiettecatte2019ultrafast}. Copyright 2019 American Chemical Society.}\label{Fig_Fit}
\end{figure}

In the examples listed above, we emphasized how $\DA$ spectra often result from a complex interplay between spectral shifts ($\DE$) and linewidth changes ($\DW$) of the exciton band(s). Additionally, these effects are supplemented by changes of the integrated absorbance of the exciton transition ($\DS$). Such changes are linked to a reduction in oscillator strength, which occurs both due to state filling/Pauli-blocking or due to screening of the Coulomb potential.\cite{steinhoff2014influence} Since $\DS$ mostly scales linearly with carrier density in low density regimes, it's dynamics are a direct metric for the carrier lifetime. As such, $\DS$ is of most interest, \textit{e.g.} in charge transfer studies or to evaluate trapping dynamics.\cite{maiti2021generating,markeev2022exciton} The sheer complexity of the resulting $\DA$ spectra made researchers devise data analysis techniques to deconvolute and quantitatively analyse the time-dependent evolution of these metrics. Throughout the literature, spectral deconvolution approaches have been used to unravel charge carrier decay in a variety of TMDs, including $\MoS$, $\ReS$, and $\mathrm{WS_2}$.\cite{cunningham2016charge,schiettecatte2021unraveling,cunningham2019resonant,zhao2020dynamics,cunningham2017photoinduced,zhou2019ultrafast,ruppert2017role,sie2017observation,trovatello2022disentangling}   
Generally, these methods rely on a curve fit to the excited-state absorbance $A$ (or reflectance $R$). We illustrate the basic principle behind this approach in Figure~\ref{Fig_Fit}a starting from a $\DA$ spectrum that results from a broadened, red-shifted and bleached Gaussian absorption profile superimposed on a background -- a shape commonly encountered in the literature.\cite{karmakar2021electron,schiettecatte2019ultrafast,zhou2019ultrafast,cunningham2016charge,ruppert2017role} First, the excited-state or nonlinear absorbance ($A$) is computed by summing the transient absorbance ($\DA$) to the ground-state or linear absorbance ($A_0$), \textit{i.e.}, $A(E,t)=A_0(E)+\Delta A(E,t)$. Both ground-state and excited-state absorbance are consequently fit to Gaussian (or where adequate Lorentzian) line-shapes $G(E)$, describing the exciton band(s), and an eventual background absorbance $C(E)$, accounting for the continuum energy transitions and/or scattering:
\begin{eqnarray}
A(E) &=& G(E) + C(E)\\
&=& h \exp ((-E-E_{X})^2/w^2) + \sum_{j}^{m} K_j E^j\label{Eq_fit2}
\end{eqnarray}
Here, $h$, $E_{X}$ and $w$ are, respectively, the amplitude, central position and linewidth of the Gaussian (see Figure~\ref{Fig_Fit}b for a graphical representation), whereas the area $S$ scales with the product of the $w$ and $h$. The background absorbance $C(E)$, on the other hand, is typically implemented as a polynomial with order $m$ and pre-factors $K_j$.

\begin{figure}[h]
\centering
\includegraphics[scale=1]{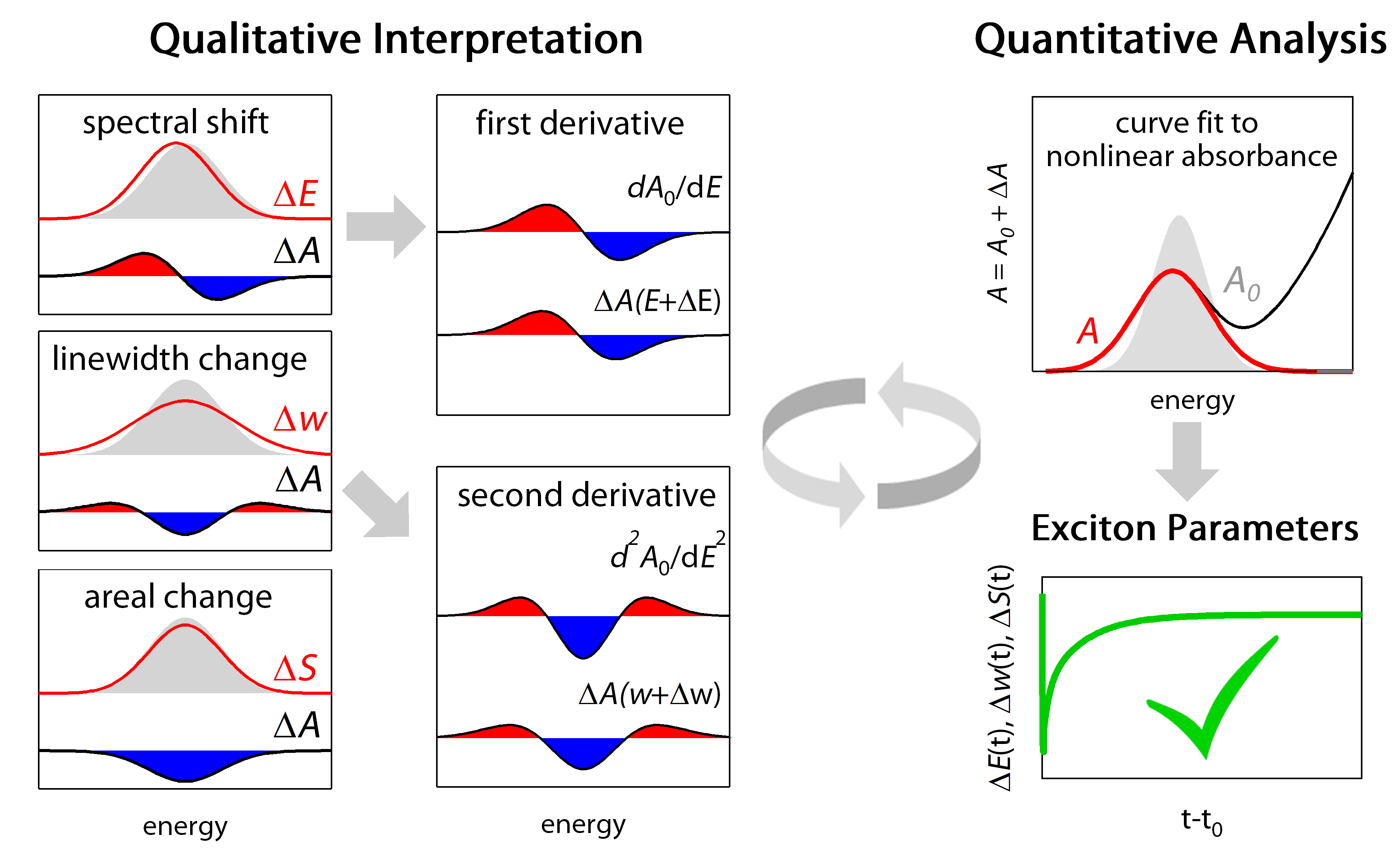}
\caption{Summary of the roadmap to interpret congested $\DA$ spectra and obtain kinetic information on intrinsic variables as exciton energy shifts $\Delta E$, linewidth changes $\Delta w$ and changes in the integrated absorbance $\Delta S$ of the exciton resonance. This toolbox includes a continuous feedback loop between (left) a qualitative interpretation based on line shapes and derivatives, and (right) a quantitative analysis based on a spectral deconvolution through rational curve fits of the full absorbance  $\A$+$\DA$. The transient absorbance $\DA$ is represented by the black curve filled in red (positive $\DA$) and blue (negative $\DA$). The filled grey curve is the linear absorbance $\A$ and the red curve is the nonlinear absorbance $A$ computed as $\A$+$\DA$.}\label{Fig_Overview}
\end{figure}

In this way, by comparing these parameters before and after photo-excitation, the transient absorbance is deconvolved into a clear set of parameters: photo-induced areal changes, spectral shifts and linewidth changes. Repeating this procedure for each pump-probe delay maps the time-dependent evolution of these fit parameters. We illustrate this method in Figure~\ref{Fig_Fit}c by fitting the $\DA$ map of $\MoS$ found in Figure~\ref{Fig_Shifts}e. From such a deconvolution, it follows that the transient absorbance resulted from a reduced oscillator strength and broadening of the $A$ and $B$ excitons and a red-shift of the entire absorbance spectrum; a result conform with a qualitative assessment based on line shapes and derivatives, see Figure~\ref{Fig_Fit}. As in particular the $A$ exciton bleach exhibited a rapid first-order decay, one could propose that non-radiative recombination proceeded through trapping of photoexcited holes in mid-gap states.\cite{schiettecatte2019ultrafast} This example shows how exciton parameters, and their dynamics, can be extracted through curve fits, backed up by insights obtained from qualitative interpretation based on line shapes and derivatives (see overview of the roadmap in Figure~\ref{Fig_Overview}). Combining such 'blind' output of a fitting algorithm with qualitative insights allows us to guide the former in the right direction. For example, based on the derivative insights we expected redshifts in the order of 20 meV. If the fitting routine would produce much smaller shifts, or even blueshifts, we had to re-adjust our numerical approach by, for example, restricting certain parameters.

\section{Conclusion}

In summary, we outlined a concise roadmap for analyzing spectrally complex transient absorbance maps based on recent literature examples. The roadmap, summarized in Figure~\ref{Fig_Overview}, offers a rational methodology to go from congested 2D time-wavelength maps to clear kinetic information on exciton parameters as the exciton energy change $\Delta E$, the linewidth change $\Delta w$ and the change of oscillator strength or the integrated change in absorbance $\Delta S$. In particular, we emphasized the synergy between an initial qualitative understanding of the transient photo-response based on line shapes and their derivatives, and a consequent quantitative spectral deconvolution backed by such insights.

\section{Acknowledgements}

P.S. acknowledges the FWO-Vlaanderen (1S40117N) and VLAIO (AIO.ONO.2020.0007.01) for funding. P.G. acknowledges support from FWO-Vlaanderen (12K8216N). Z.H. acknowledges the Research Foundation Flanders (research projects 17006602 and G0F0920N) and Ghent University (GOA no. 01G01513) for funding.

\bibliography{bibfile}

\end{document}